\documentclass[a4paper,11pt]{article}
\usepackage{pos}

\title{Angular analyses of rare decays at the LHC}

\author*[a]{Biplab Dey, on behalf of the LHCb collaboration}

\affiliation[a]{Eotvos University,\\
  Budapest, Hungary}

\emailAdd{biplab.dey@cern.ch}

\abstract{Loop-suppressed penguin $b\to s$ transitions are sensitive to heavy New Physics particles propagating inside the loops. Thanks to the large sample sizes from the LHC, we are able to perform multidimensional angular analyses that are sensitive to interferences between the Standard Model and New Physics terms. This article surveys the latest results, primarily from LHCb, on $b\to s\mup\mun$ electroweak and $b\to s\gamma$ radiative penguins.}

\FullConference{
16th International Conference on Heavy Quarks and Leptons (HQL2023)\\
28 November-2 December 2023\\
TIFR, Mumbai, Maharashtra, India\\}

\input{aliases}
\begin{document}
\maketitle

\section{Introduction and theory}

\begin{figure}
    \centering
    \includegraphics[width=\textwidth]{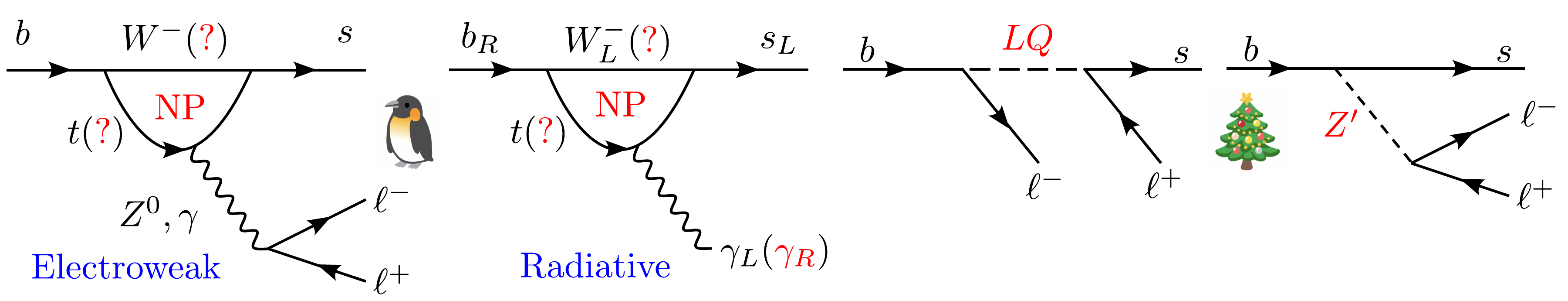}
    \caption{Flavor-changing neutral currents occur only at the loop-level in the SM but can be enhanced by NP effects, both in loop (penguin) and tree-level diagrams.}
    \label{fig:intro}
\end{figure}

In the Standard Model (SM), the flavor-changing neutral current process $b\to s$ is forbidden at the tree-level and proceeds only via loop-suppressed diagrams as shown in Fig.~\ref{fig:intro}. These provide excellent avenues to probe New Physics (NP) contributions that can enter either in loop- or tree-level processes such as via Leptoquarks (LQ) or heavy $Z'$ boson, as shown in Fig.~\ref{fig:intro}. This article focuses on the electroweak (EWP) and radiative (Rad) penguin diagrams, but gluonic penguins can also be an important NP source. A convenient theoretical formalism to study such decays is to regard the SM as a low energy effective field theory containing dimension $d\leq 4$ local operators from renormalizability requirements. Higher dimensional operators can be added with an appropriate cutoff scale $\Lambda$, as
\begin{align}
\mathcal{L}_{\rm eff}(x) =  \displaystyle \mathcal{L}_{\rm SM}(x) + \sum_{d>4} \frac{C_i}{\Lz^{d-4}} \mathcal{O}_i^{(d)} (x)
\end{align}
whereby the NP amplitudes have $(E/\Lz)^{d-4}$ behavior in the energy, $E$: divergent at high energies, but suppressed at $E\ll \Lz$. Most relevant for rare $b\to s$ decays are $d=6$ operators that yield $\mathcal{A}_{\rm eff} \sim C^{\rm SM}/m^2_W + C^{\rm NP}/\Lz^2_{\rm NP}$. The basis comprises 10 operators~\cite{Altmannshofer:2008dz}: $\mathcal{O}_{1,2}$ (4-quark tree), $\mathcal{O}_{3-6}$ (4-quark penguins) and $\mathcal{O}_{8}$ (gluonic penguin) that are suppressed for the EWP/Rad modes. The dominant left-handed contributions are from the electromagnetic dipole and weak vector (axialvector) operators
\begin{align}
\mathcal{O}_{7\gamma} = \frac{e}{16\pi^2} m_b (\overline{s} \sigma_{\mu\nu} P_R b) F^{\mu \nu},\;\;\; \mathcal{O}_{9V (10A)} = \frac{e^2}{16\pi^2}  (\overline{s} \gamma_\mu P_L b) (\overline{\ell} \gamma^\mu (\gamma_5)\ell).  
\label{eq:effops}
\end{align}
The corresponding right-handed (quark side) operators are suppressed in the SM, but can be enhanced in NP scenarios. The dimensionless couplings (Wilson coefficients) associated with the operators in Eq.~\ref{eq:effops} encode the short distance physics. They are calculated at the $m_W$ scale by integrating out the heavy degrees of freedom from the full theory and evolving to the $m_b$ scale using renormalization group equations. The total amplitudes $\mathcal{A}(i\to f) = \displaystyle C_n(m_b) \langle f| \mathcal{O}_n(m_b) |i \rangle_{\rm had}$ also contains the long-distance physics (QCD/hadronization) which mostly comes from {\em local} form-factors (FFs) that are computed from lattice QCD and other theory tools, but can get important non-local contributions (rescattering, charm loops) that are hard to estimate theoretically.

\begin{figure}
    \centering
    \includegraphics[width=\textwidth]{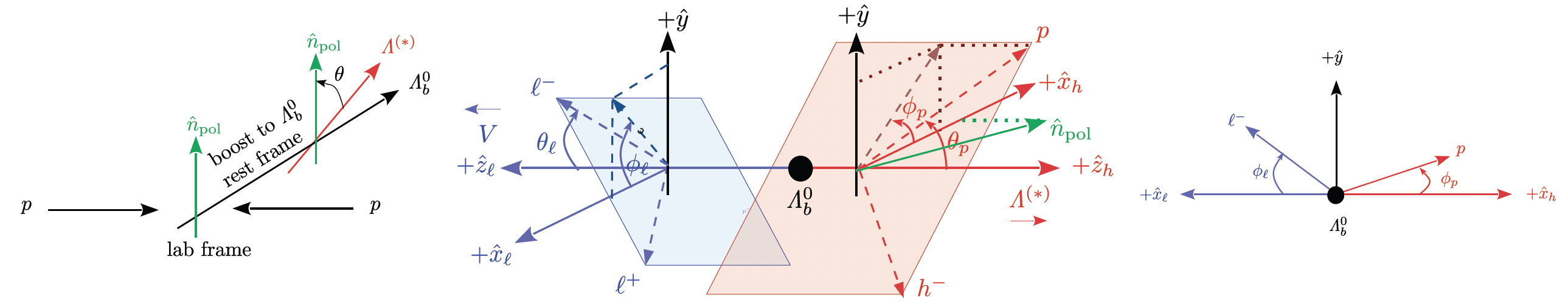}
    \caption{The four angular variables for the decay of a polarized $\vec{\Lb}$ decaying to a 4-body final state, $p\Km\ellp\ellm$: $\thetal$ ($\theta_p$) is the $\ellm$ (proton) helicity angle. The beam direction, $\Lz^{\ast}$ flight direction and the $\Lb$ polarization defines a coordinate system relative to which the dilepton and dihadron azimuthal angles can be defined as $\phi_\ell$ and $\phi_p$.}
    \label{fig:ang_vars}
\end{figure}

Thanks to the large $b\overline{b}$ samples available at the LHC, a comprehensive effort on multidimensional angular analyses in $b\to s\ellp\ellm$ and $b\to s\vec{\gamma}$ is ongoing. These offer a rich set of angular observables sensitive to $\Delta C_i\equiv C_i^{\rm SM} -  C_i^{\rm NP}$. The thrust has been to identify and probe ``theoretically clean observables'' with reduced dependence on the QCD contributions that often form the largest theory uncertainties. The angles for a typical 4-body final state decay is shown in Fig.~\ref{fig:ang_vars} for $\vec{\Lb}\to p\Km\ellp\ellm$; the two other kinematic variables being $\qsq\equiv m^2_{\ellp\ellm}$ and $k^2\equiv m^2_{pK}$. For an unpolarized parent $\Lb$ (or spin-0 $B_{(s)}$ mesons), $\phi_\ell$ is set to 0 and $\chi= \phi_p$ is the single azimuthal angle between the dilepton and dihadron decay planes. Equivalent variables apply for $B\to K\pi \ellp\ellm$ and $\Bs^0\to \Kp\Km\ellp\ellm$.

\section{Electroweak penguins}
\subsection{The golden channel: $\Bz\to K^{\ast 0} \ellp\ellm$}

\begin{figure}[h]
    \centering
    \includegraphics[width=0.85\textwidth]{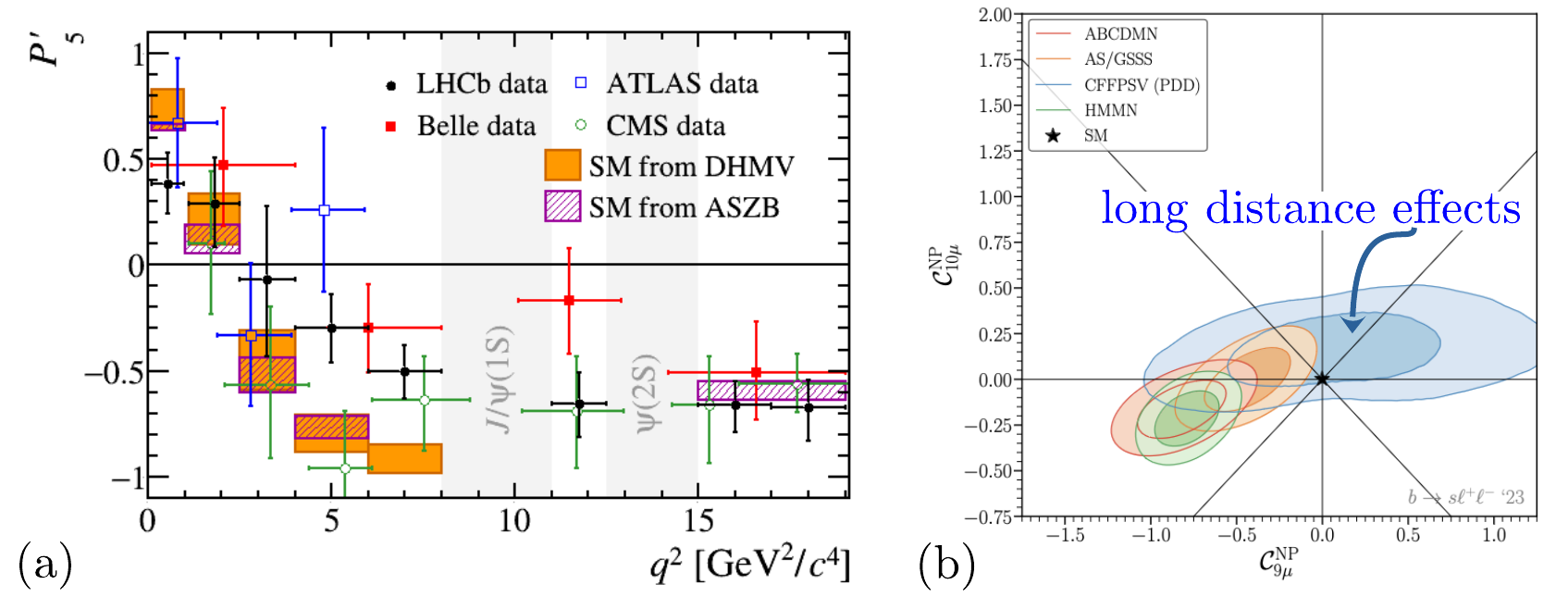}
    \caption{$\Bz\to K^{\ast 0}\mup\mun$: (a) world-average of the $P'_5$ measurements compared to theory (filled/hatched boxes). The situation is representative after publication of the LHCb Run~1 results. (b) effect on $C^{\rm NP}_{9,10}$ from a recent global fit in Ref.~\cite{Capdevila:2023yhq}. }
    \label{fig:pos_p5prime}
\end{figure}

$\Bz\to K^{\ast 0} \ellp\ellm$ has been a gold-plated channel~\cite{Altmannshofer:2008dz} since the $B$-factory era due to the relatively narrow $K^{\ast 0}(892)$ resonance. Especially in the low $\qsq$ regime, where the recoiling $K^{\ast 0}$ has a large $\gamma$-factor in the parent $\Bz$ frame, QCD sum rules on the light-cone (LCSR) affords control over the FF calculations. However, with the BaBar/Belle statistics, only 1-dimensional angular analyses in either the lepton or hadron helicity angles ($\cos \thetal$ or $\cos\theta_K$) were possible~\cite{BaBar:2012mrf}. Full 3-dimensional analysis in $d\thetal d\theta_K d\chi$ in $\qsq$ bins was possible only with the advent of LHCb. For instance, while the full BaBar dataset had $\mathcal{O}(50)$ $\Bz\to K^{\ast 0} (\to \Kp\pim) \mup\mun$ events, the existing Run1+2 LHCb dataset already includes $\mathcal{O}(10^4)$ clean signal events for this muonic mode. On the other hand, $B$-factories, including Belle~II, have complementary advantages, with better reconstruction for the $\piz$ isospin modes as well the dielectron channels.

\begin{figure}[h]
    \centering
    \includegraphics[width=\textwidth]{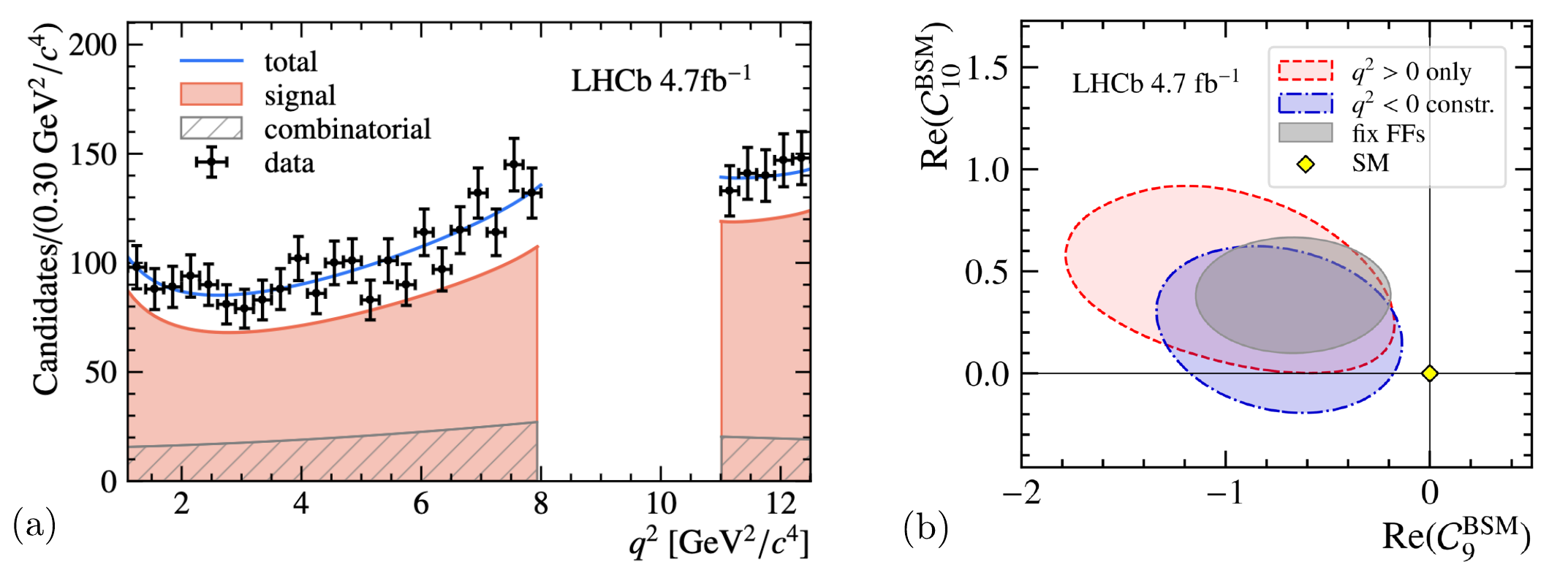}
    \caption{$\Bz\to K^{\ast 0}\mup\mun$ long-distance contribution fits~\cite{LHCb:2023gpo,LHCb:2023gel}: (a) $\qsq$ projections; (b) effect on $Re(C^{\rm NP}_{9,10})$.}
    \label{fig:h_lambda_results_pos}
\end{figure}

An important result from the muonic analyses at LHCb is tension with SM predictions in the angular observable $P'_5$~\cite{LHCb:2020lmf} (see Fig.~\ref{fig:pos_p5prime}a). Similar tensions have also been seen in $B^+\to K^{\ast +}\mup\mun$~\cite{LHCb:2020gog} and $\Bs^0\to\phi\mup\mun$~\cite{LHCb:2021xxq}. A related tension is observed in the overall branching fractions in several $b\to s\mup\mun$ modes which tend to consistently lie lower than the SM predictions~\cite{LHCb:2021zwz,LHCb:2015tgy,LHCb:2014cxe}. Competitive results have also come from ATLAS/CMS~\cite{ATLAS:2018gqc,CMS:2017rzx,CMS:2020oqb} where the advantage is higher overall luminosity, but the disadvantage is the limited $B$-physics trigger bandwidth and lack of a RICH detector for $\Kp/\pip/p$ separation. Numerous global fits with different data subsets, statistical methods and theory priors have been performed~\cite{Capdevila:2023yhq} pointing to a preferred negative $C^{\rm NP}_9$. The major point of contention, however, has been the effect of non-factorizable long-distance contributions due to soft+hard gluons from charm-loops that can mimic NP effects (see Fig.~\ref{fig:pos_p5prime}b). To constrain the non-factorizable part in a data-driven fashion, LHCb has performed  an unbinned angular analysis~\cite{LHCb:2023gpo,LHCb:2023gel} using the same dataset as in Ref.~\cite{LHCb:2020gog}. The underlying transversity amplitudes are
\begin{align}
\mathcal{A}^{L,R}_{\tiny \lambda= 0,\parallel,\perp} =& N_\lambda\left\{ \textcolor{black}{ \left[(C_9 \pm C'_9)\mp (C_{10} \pm C'_{10})\right] }\mathcal{F}_\lambda + \frac{2m_b M_B}{q^2}  \left[  (C_7 \pm C'_7) \mathcal{F}^T_\lambda  - 16 \pi^2 \frac{M_B}{m_b} \textcolor{black}{\mathcal{H}_\lambda}  \right]  \right\}, 
\end{align}
where $\mathcal{F}_\lambda$ are the usual local FFs (taken from LCSR and lattice QCD) and $\mathcal{H}_\lambda$ are the new non-factorizable part which are extracted from a $\qsq$-dependent parameterization. The values of $C_i^{\rm SM}$ are taken from theory, while allowing for $C_{9,10}^{\rm NP}$ to be floated. The results of the fit are shown in Fig.~\ref{fig:h_lambda_results_pos}. Good consistency is found in the extracted binned observables compared to Ref.~\cite{LHCb:2020gog}. The overall tension with the SM is reduced to $\sim 1.8\sigma$ in $C_9$, and $\sim 1.4\sigma$ in global fits. The full Run~1+2 analysis and more precise theory FFs will improve upon these results. 

\subsection{Access to tensor states in $B\to X_s\mup\mun$}

\begin{figure}
    \centering
    \includegraphics[width=\textwidth]{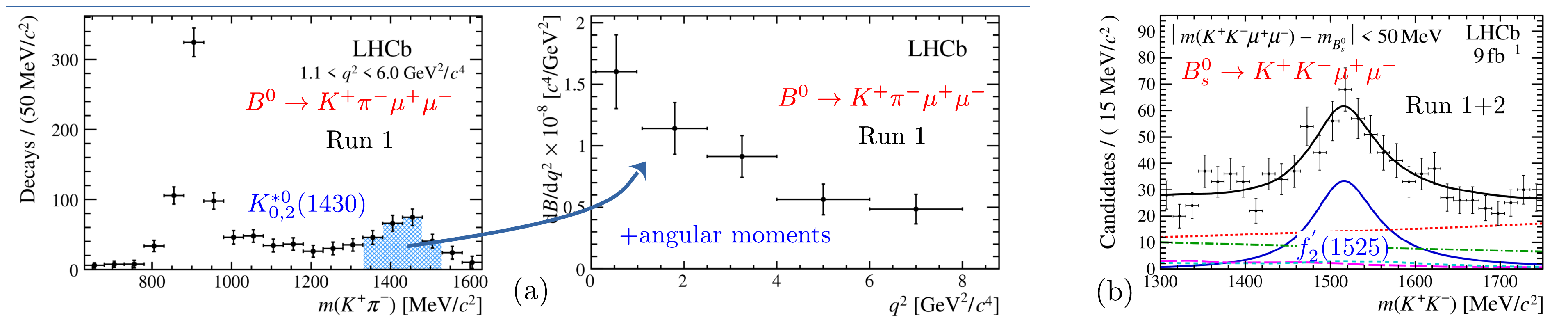}
    \caption{Tensor states in (a) $B^0\to \Kp\pim\mup\mun$, (b) $B^0_s\to \Kp\Km\mup\mun$ decays.}
    \label{fig:excited_hh_pos}
\end{figure}

While most of the theory and experimental investigations have focused on the ground state vector states $K^{\ast}(892)$ and $\phi(1020)$ in $B$-meson decays, LHCb has also probed the excited $K^\ast_{0,2}(1430)\to \Kp\pim$~\cite{LHCb:2016eyu} and $f'_2(1525)\to \Kp\Km$~\cite{LHCb:2021zwz}, including an angular moments analysis~\cite{Dey:2015rqa} for the former, to separate the $S$-, $P$- and $D$-wave $K\pi$ states. The results are shown in Fig.~\ref{fig:excited_hh_pos}. The theory interpretation however will require reliable FFs for $B$ decays to these excited states.

\subsection{Angular analysis of $B\to K\mup\mun$}

\begin{figure}[h!]
    \centering
    \includegraphics[width=0.9\textwidth]{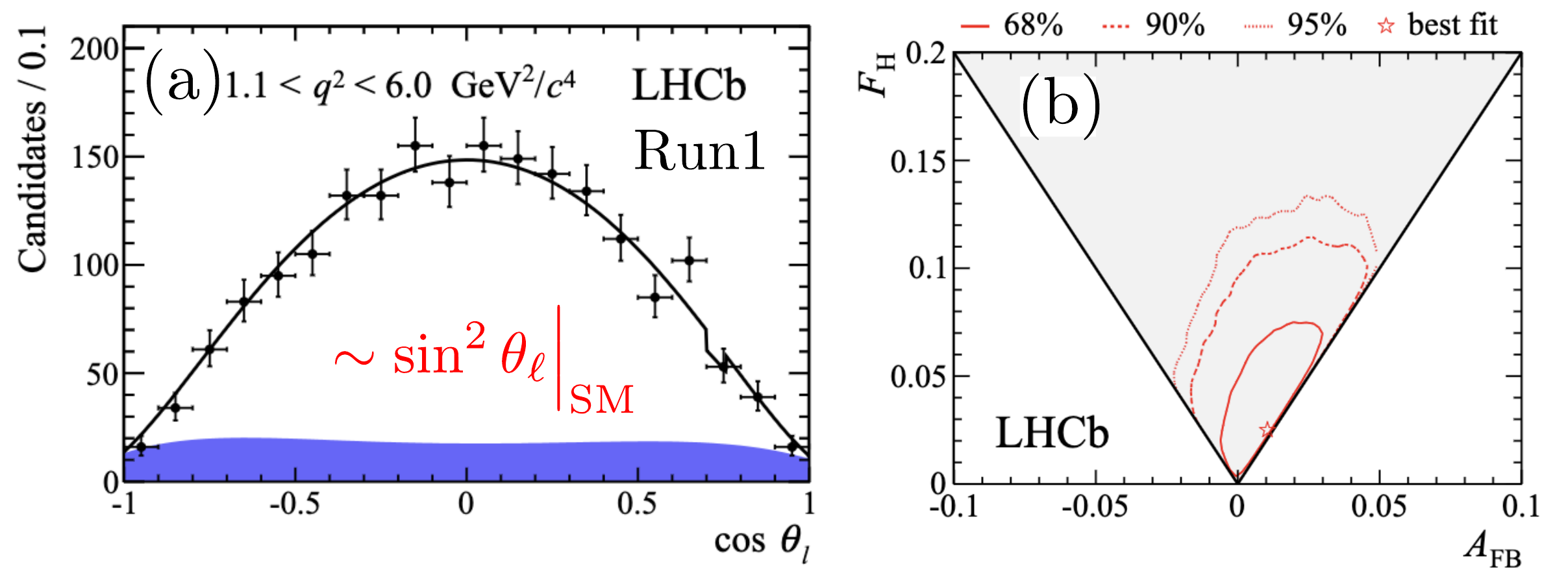}
    \caption{Charged and neutral $B\to K\mup\mun$ angular analysis~\cite{LHCb:2014auh}: (a) projection in $\cos\thetal$ showing the expected $\sin^2\thetal$ distribution. (b) extracted $A_{\rm FB}$ and $F_{\rm H}$ with uncertainty contours.}
    \label{fig:kmm_pos}
\end{figure}

For the 3-body final state $B\to K\mup\mun$, only the lepton helicity angle, $\thetal$, can be defined and the SM predicts an almost pure $\sin ^2\thetal$ distribution save for small effects due to the muon mass. The angular distribution is be sensitive to new scalar and tensor operators via the new terms $A_{\rm FB}$ and $F_{\rm H}$:
\begin{align}
\label{eqn:kmm}
    d\Gamma/d\cos\thetal \propto \frac{3}{4} (1-\textcolor{black}{F_{\rm H}}) \sin^2\theta_\ell +  \frac{1}{2} \textcolor{black}{F_{\rm H}}  + \textcolor{black}{A_{\rm FB}} \cos\theta_\ell.
\end{align}
Figure~\ref{fig:kmm_pos} shows the results of the fit to Run~1 data for both $B^{+,0}$~\cite{LHCb:2014auh}, subject to the constraints $|A_{\rm FB}| \leq F_{\rm H}/2$, $0\leq F_{\rm H} \leq 3$, such that the rate in Eq.~\ref{eqn:kmm} is positive. The extracted $A_{\rm FB}$ and $F_{\rm H}$ in different $\qsq$ bins are also consistent with SM.

\subsection{$\Lb\to \Lz\mup\mun$ moments analysis}

\begin{figure}[h!]
    \centering
    \includegraphics[width=0.9\textwidth]{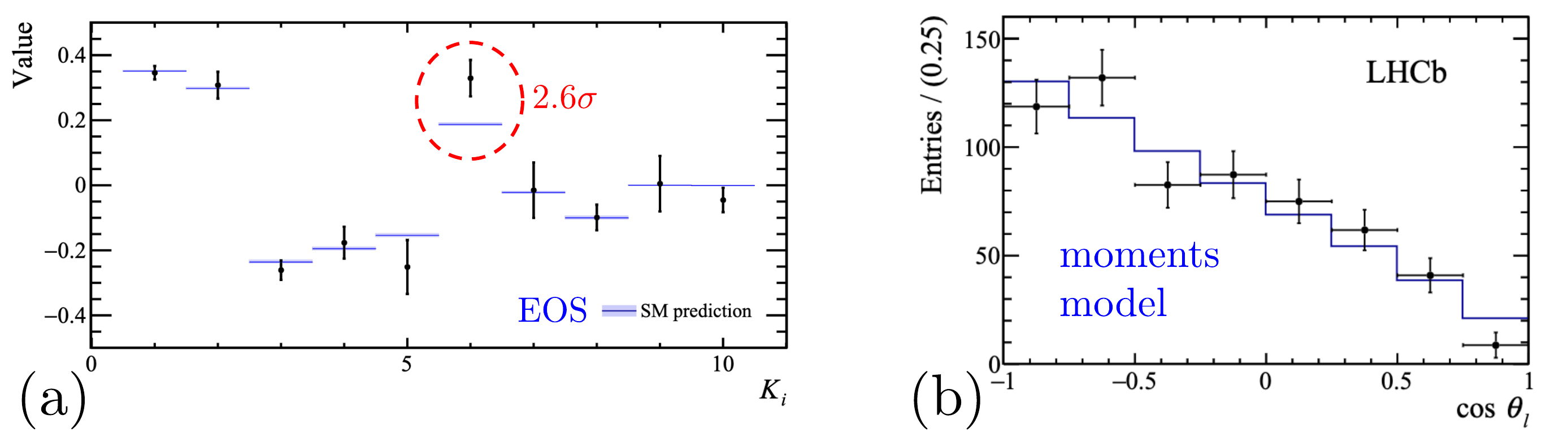}
    \caption{$\Lb\to \Lz^0 \mup\mun$ angular analysis in $\qsq\in [15,20]$~GeV$^2$~\cite{LHCb:2018jna}: (a) first 10 moments from Eq.~\ref{eqn:lz_mom}; (b) comparison of the background-subtracted data and angular moments model.}
    \label{fig:lb2lzmm_mom_pos}
\end{figure}

LHCb is unique among the $B$-factories to have access to all $b$-hadron species, including large samples of $\Lb$ baryons. Therefore it is possible to probe $b\to s$ penguin decays in the baryonic sector as well. For $\Lb\to \Lz^0$ transition, the narrow $\Lz^0$ state allows lattice QCD calculations~\cite{Detmold:2016pkz} for the FFs. For a given value of $\qsq$, the decay rate for polarized $\vec{\Lb}$ and $\vec{\Lz^0}$ depends on five angles, $\vec{\Omega}= \{\theta, \thetal, \phi_\ell, \theta_h, \phi_h \}$ (see Fig.~\ref{fig:ang_vars}) and is expanded in an orthonormal angular basis as
\begin{align}
\label{eqn:lz_mom}
    \frac{d^5\Gamma(\qsq)}{d \vec{\Omega}} = \frac{3}{32\pi^2} \sum_{i=1}^{34} K_i(\qsq) f_i(\vec{\Omega}).
\end{align}
The $K_i$ moments can be related to more familiar observables such as forward-backward asymmetries in the angles: $A^\ell_{\rm FB} = [\frac{3}{2} K_3] (\to \cos\thetal)$, $A^h_{\rm FB} = [K_4 +\frac{1}{2} K_5] (\to \cos\theta_h)$, $A^{\ell h}_{\rm FB} = [\frac{3}{4} K_6] (\to \cos\thetal \cos\theta_h)$. If the $\Lb$ is unpolarized, only the first 10 moments are non-zero~\cite{eos}. Due to the long-lived nature of the $\Lz^0$, its reconstruction in LHCb is somewhat non-trivial. At low $\qsq$, the Run~1 analysis~\cite{LHCb:2015tgy} found very few events. Therefore the analysis using Run~1 + partial Run~2 (collected between 2011-16)~\cite{LHCb:2018jna} focused on the high-$\qsq$ region, $\qsq \in [15,20]$~GeV$^2$. The results are shown in Fig.~\ref{fig:lb2lzmm_mom_pos}. The SM predictions are taken from the {\tt EOS}~\cite{eos} software package and show a slight tension with the data in the $K_6$ observable.

\subsection{Differential cross-sections for $\Lb\to \PLambda(1520) \mup\mun$}
\label{sec:pkmm}
\begin{figure}[h!]
    \centering
    \includegraphics[width=0.9\textwidth]{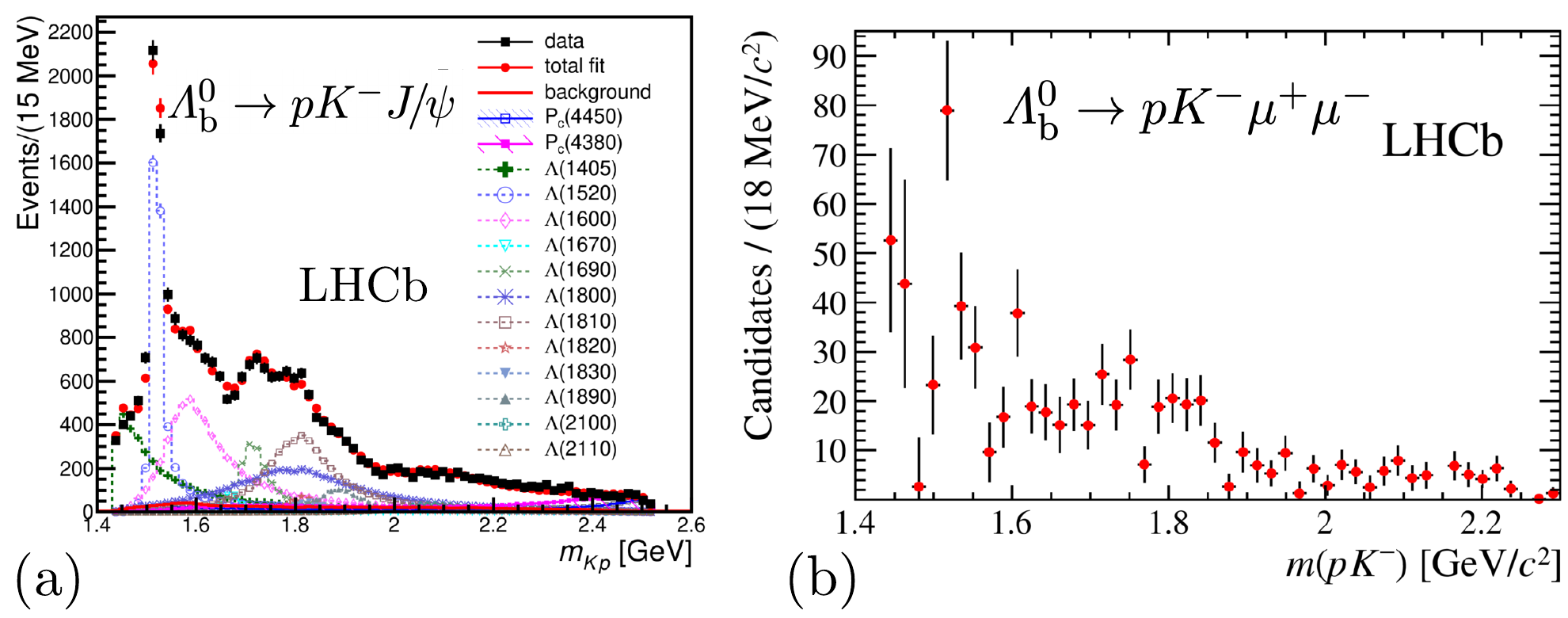}
    \caption{$m(p\Km)$ spectrum from Run~1 $\Lb\to p\Km \mup\mun$ LHCb data: (a) at the $\jpsi$ resonance~\cite{LHCb:2015yax}; (b) in the non-resonant region~\cite{LHCb:2017slr}.}
    \label{fig:pkmm_pos}
\end{figure}

\begin{figure}
    \centering
    \includegraphics[width=\textwidth]{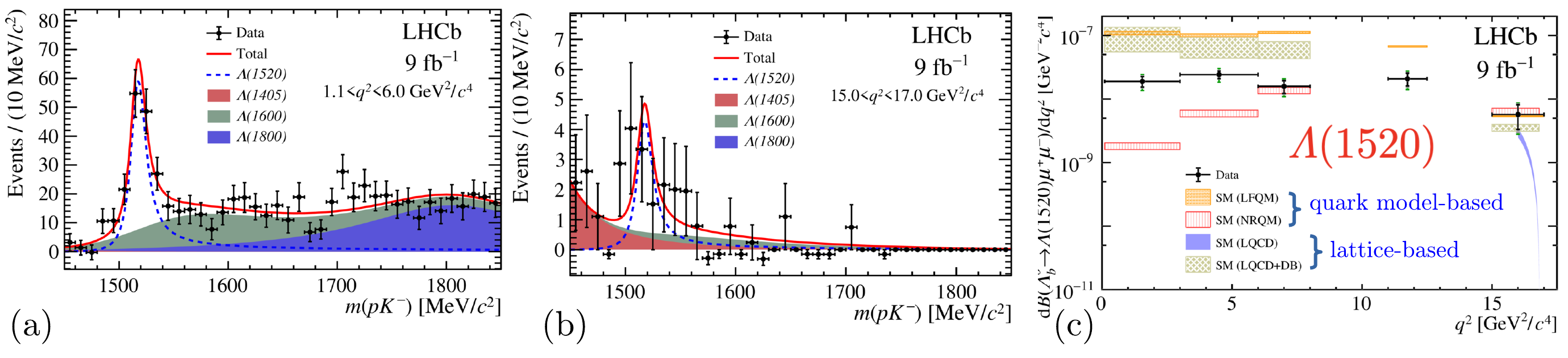}
    \caption{$\Lb\to \PLambda(1520) \mup\mun$~\cite{LHCb:2023ptw}: (a) low $\qsq$, (b) high $\qsq$, (c) $d\Gamma/d\qsq$.}
    \label{fig:pkmm_results_pos}
\end{figure}

The first observation of $\Lb\to p\Km\mup\mun$ decay using LHCb Run~1 data~\cite{LHCb:2017slr} demonstrated a rich $m(pK)$ spectrum. The comparison between the resonant and non-resonant spectra is shown in Fig.~\ref{fig:pkmm_pos}. Employing the full Run~1+2 datasets, LHCb has examined the $m(p\Km)$ region around the narrow $\Lambda(1520)$ resonance. Integrating over the angles and including the resonances $\Lz(1405) (\frac{1}{2}^-)$, $\Lz(1520) (\frac{3}{2}^-)$, $\Lz(1600) (\frac{1}{2}^+)$ and $\Lz(1800) (\frac{1}{2}^-)$, 1-dimensional fits are performed in $m(p\Km)$. Differential cross-sections for the $\Lb\to \PLambda(1520) \mup\mun$ decay are also provided~\cite{LHCb:2023ptw} that show large discrepancies with theory calculations in the low $\qsq$ region (see Fig.~\ref{fig:pkmm_results_pos}).

\section{Radiative penguins}

In the SM, due to the left-handed nature of the weak interaction, the photon from a $b\to s\gamma$ is almost purely left-handed; the right-handed component is suppressed ($C'_7\sim \frac{m_s}{m_b} C_7$) and is a sensitive NP probe. However, one needs angular analyses to extract the interferences.  LHCb probes this in several ways: very low $\qsq$ angular analysis of $\Bz\to K^{\ast 0} e^+ e^-$~\cite{LHCb:2020dof} and $\B^0_s\to \phi e^+ e^-$; angular analysis of $\Bp\to \Kp\pip\pim\gamma$~\cite{LHCb:2014vnw},; time-dependent CP violation of $B^0_s\to \phi\gamma$~\cite{LHCb:2016oeh} and $B^0\to \KS\pip\pim\gamma$; angular analyses of $\Lb \to \Lz^{0(\ast) }\gamma$~\cite{LHCb:2021byf}.

\subsection{Angular analysis of $\Lb\to \Lz^0\gamma$}

\begin{figure}[h!]
    \centering
    \includegraphics[width=0.9\textwidth]{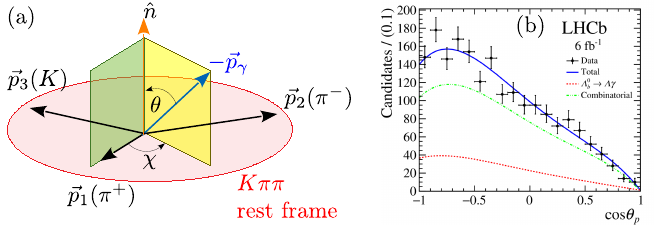}
    \caption{(a) If the hadronic system has a 3-body decay, the normal to the plane defines a preferred direction to extract the photon polarization from the up-down asymmetry in $\cos\theta$~\cite{LHCb:2014vnw}. (b) An exception is the 2-body decay of $\vec{\Lz}\to p\pim$ where the measurable polarization of the $\Lz^0$ provides the preferred direction.}
    \label{fig:phpol_udasym}
\end{figure}

To measure the photon polarization in $H_b \to H_s \vec{\gamma}$, the hadronic system must undergo a 3-body decay, so that the normal to the plane defines a preferred direction and the up-down asymmetry is proportional to the photon polarization, $\lambda_\gamma$. For example, this has been utilized in $\Bp\to \Kp\pip\pim\gamma$~\cite{LHCb:2014vnw}  as shown in Fig~\ref{fig:phpol_udasym}a. In such cases, due to poor knowledge of the resonant structures in the $H_s$ system and thereby the hadronic current $\mathcal{J}_\mu^{\rm had}$, the proportionality factor remains unknown and $\lambda_\gamma$ can still not be extracted out. An exception is the 2-body decay of  $\vec{\Lz^0}\to p\pim$, where the self-analyzing nature of the $\Lz^0$ polarization provides the preferred direction. Moreover, the differential rate is $d\Gamma/d\cos\theta_p \propto (1 - \alpha_\Lambda \lambda_\gamma \cos\theta)$ where the $\Lz^0$ decay asymmetry parameter $\alpha_\Lambda$ is known quite precisely~\cite{Workman:2022ynf} and thereby $\lambda_\gamma$ can be extracted from a fit to the $\cos\theta_p$ slope, as shown in Fig~\ref{fig:phpol_udasym}b. Experimentally, $\Lb\to \Lz^0\gamma$ is challenging due to lack of a reconstructible secondary vertex for the $\Lb$ decay. Employing dedicated reconstruction to reject the high background, the first observation with Run~2 2016 data was reported in Ref.~\cite{LHCb:2019wwi} and the polarization measurement with full Run~2 was reported in Ref.~\cite{LHCb:2021byf}. The measured value of $\alpha_\gamma = 0.82^{+0.17}_{-0..26} {\rm (stat.)}^{+0.04}_{-0.13}{\rm (syst.)}$ is compatible with the SM expectation of 1 and from global $C_7^{(')}$ fits, reducing a 4-fold ambiguity in the $C_7^{\rm NP}$ phase to a 2-fold ambiguity.

\subsection{Angular analysis of $\Lb\to p\Km\gamma$}

\begin{figure}[h!]
    \centering
    \includegraphics[width=\textwidth]{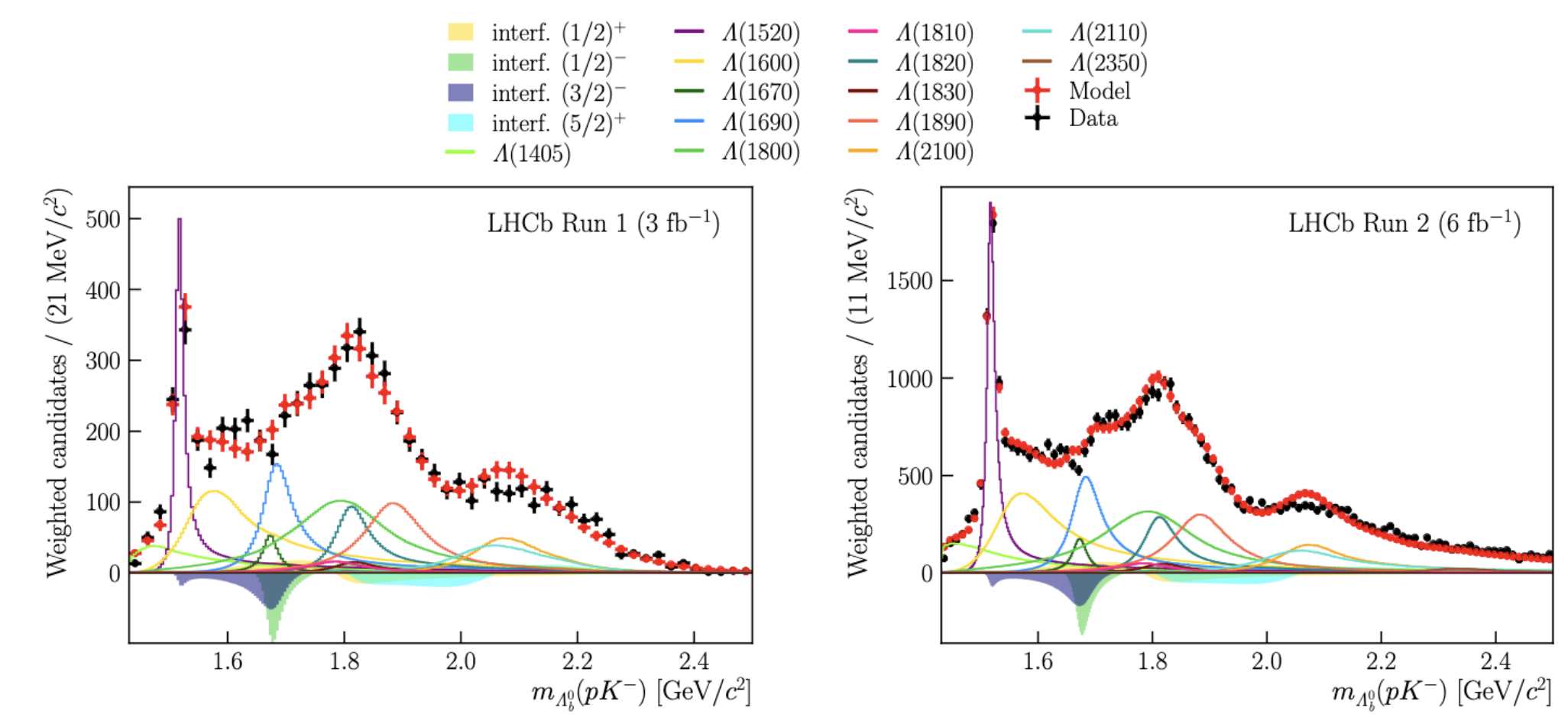}
    \caption{Fit projections for $\Lb\to p\Km\gamma$ in the $m(pK)$ varaiable. The shaded areas depict interferences between $\Lz^\ast$ states with the same spin-paiity.}
    \label{fig:pkg}
\end{figure}

As for the electroweak penguin case in Sec.~\ref{sec:pkmm},  for the radiative $\Lb\to \Lz^\ast(p\Km)\gamma$ decay, the $[p\Km]$ system is dominated by a large number of overlapping resonances. Employing the full Run1+2 data, LHCb has performed a 2-dimensional angular analysis of this mode in $\{m(pK), \cos\theta_p\}$. All well-established $\Lz^\ast$ states~\cite{Workman:2022ynf} are included in an isobar model with Breit-Wigner lineshapes incorporating mass-dependent widths. The only exception is the $\Lz(1405)$ resonance where a two-pole Flatte-form is used. To reduce the large set of parameters, the maximum orbital angular momentum of the $\Lz^\ast \to p\Km$ decay is taken to be $L=3$. The result from such a  {\em reduced} model is shown in Fig.~\ref{fig:pkg}. The fit fractions and interference fit fractions are reported from the amplitude analysis.

\section{Summary and outlook}

The large existing Run~1+2 data samples at the LHC has already allowed the hitherto (pre-LHC)  ``rare'' $b\to s$ decays to be probed in an unprecedented fashion. This will continue into the LHCb upgrade era, with Run~3 having already commenced. Multi-dimensional angular analyses in both the electroweak and radiative penguin sectors have given rise to surprising tensions. A large and mature angular analysis effort exists at LHCb which will continue to be evolve. The thrust will be to include time-dependent CP violation~\cite{LHCb:2019vks} in more rare decay angular analyses, as well as to probe the further Cabibbo-suppressed $b\to d$ transitions~\cite{LHCb:2017lpt,LHCb:2015hsa}.

\bibliographystyle{JHEP}
\bibliography{main}

\end{document}